# Modeling of End-Use Energy Profile: An Appliance-Data-Driven Stochastic Approach

Zhaoyi Kang[1], Ming Jin[1], Costas J. Spanos[1]
[1]Dept. of Electrical Engineering & Computer Sciences, University of California, Berkeley, Berkeley CA, USA
[1]{kangzy, jinming, spanos}@berkeley.edu

*Abstract*—**In this paper, the modeling of building end-use energy profile is comprehensively investigated.** *Top-down* **and** *Bottom-up* **approaches are discussed with a focus on the latter for better integration with occupant information. Compared to the Time-Of-Use (TOU) data used in previous** *Bottom-up* **models, this work utilizes high frequency sampled appliance power consumption data from wireless sensor network, and hence builds an appliance-data-driven probability based end-use energy profile model. ON/OFF probabilities of appliances are used in this model, to build a non-homogeneous Markov Chain, compared to the duration statistics based model that is widely used in other works. The simulation results show the capability of the model to capture the diversity and variability of different categories of end-use appliance energy profile, which can further help on the design of a modern robust building power system.**

## I. Introduction

Buildings account for more than 40% of the total power consumption in the US, and can play a critical role in addressing the current energy and climate issues **[1]**. Significant effort has been invested in this topic, from benchmarking, to control and monitoring. In this paper, we will discuss the modeling of end-use energy profile of the commercial building power system.

The modeling of end-use energy profile is an important task, and is of particular interests especially in recent years because of the following reasons. Nowadays, building energy usually depends greatly on occupant behavior, especially at fine-grained metering level, such as plug-in loads, user-controlled lighting, user-adjusted HVAC, etc. **[2]**, which brings about significant amount of diversity and fluctuation. End-use profile is believed to be able to capture, quantify and predict those variability and complicated relationships.

Moreover, as people endeavor to integrate renewable energy resources to traditional building power system, and the wide adoption of energy-efficient appliances and policies, we need accurate and robust models to understand the feasibility of such schemes and to evaluate the effects of such innovations.

Last but not least, as an important potential input of building energy & indoor climate simulation software, end-use energy profile can be widely used in early-stage building environmental design and energy system planning.

This paper is organized as follows. In Section II, a literature review is given. In Section III, the data collection and processing methods in this work is described. In Section IV, the key modules in the model are illustrated and investigated. In Section V, we run simulation and discuss the results. Finally, conclusions are drawn and discussed in Section VI.

## II. Literature Review

The models of end-use energy profile can be divided into two categories, the "*Top-down* approach" and the "*Bottom-up* approach", with reference to the hierarchical structure of data inside the whole system **[3]**, as illustrated in **Fig 1 [4]**.

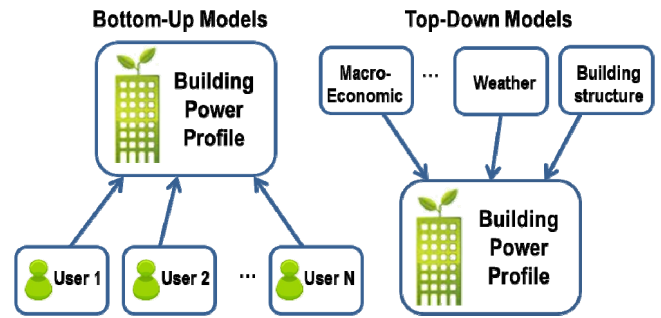

Fig. 1. Two types of approaches: Top-down and Bottom-up

The *Top-down* approach treats building as a black-box, without considering the detailed occupant-oriented behavior inside. Usually, the diversity or variability of the end-use energy profile is captured as a linear model based on *macro*-scale extraneous variables such as macroeconomic indicators (gross domestic product (GDP), income, price rate), climatic conditions, building construction, etc. **[4]**. The parameters of the statistical model are estimated from training data and the end-use profile of a new building can be extrapolated.

The *Bottom-up* approach, on the contrary, takes into consideration the contribution of each individual sector. Specifically, the occupant-oriented energy consumption is included, and the variability is captured either as a statistical model of the users based on *macro*-scale information, or stochastic sequences of the user patterns. The parameters of the models are estimated from a group of building energy consumption data or Time-Of-Use (TOU) survey data.

*Bottom-up* approaches are more recent and attracting attention, because of the following reasons:

- The building energy performance becomes more and more sensitive to occupant behaviors. The occupant-dependent variability is better captured by *Bottom-up* approach, whereas *Top-down* approach does not typically have the flexibility to model that.

- Detailed effect and feasibility analysis of particular innovations, new policies, or social games can be better investigated by the *Bottom-up* approach.

- The *Bottom-up* approaches better adapt to the changes in the building infrastructure or technology, while the *Top-down* approach relies a lot on historical data.

One of the earliest works of *Bottom-up* approach is by A. Capasso *et al.* **[5]**. They use availability probability to model presence of each member in a house, and activity probability to model presence of each activity. The probabilities are learned from TOU data. Together with duration statistics obtained from prior knowledge, power stream can be generated by Monte Carlo (MC) simulation. In **[6]**, TOU data is also used, and nine synthetic activity patterns are defined. Non-homogeneous Markov Chain is used to model the ON/OFF of activities. Duration and ON events are sampled randomly. In **[7]**, activity probability is trained from TOU data and other extraneous data, so that is non-homogeneous. In **[8]**, effort is put purely in estimating activity probability patterns based on TOU survey and duration statistics.

The existing methods that employ the *Bottom-up* approach provide great amount of insights into the variability of end-use energy. However, there are still several issues in the existing methods that need to be addressed:

- Previous works are mostly based on TOU data to obtain activity probability model, and then convert the activity pattern to appliance pattern. Since most of the conversion is not rigorously investigated, this method is sometimes problematic.

- Cross-correlations among appliances are not captured because of the conversion mentioned before. A random Markov Chain model would under-estimate the peak demand. Moreover, most previous works put emphasis on modeling *shared activities*, whereas validation of those models is difficult.

- In commercial buildings, variation among buildings is not of significant interest since the infrastructure can vary a lot. However, the variation caused by occupant fluctuation becomes especially important.

In this work, we will directly estimate probability patterns of appliances in commercial building, thanks to the large-scale wireless sensor network and distributed data storage system. We use a method combining appliance ON/OFF probability and the intra-building variation can be captured.

### III. DATA PROCESSING

*A. Building Profile*

Our experimental space is in 406 Cory Hall at University of California Berkeley, an office with about 25 occupants. Depending on the sets of appliances that each user owns, we can divide the users into six categories: A) 1 laptop, 1 monitor, 1 desktop; B) 1 desktop, 1 monitor; C) 1 laptop, 1 monitor; D) 1 laptop, 2 monitors; E) 2 laptops, 1 monitor; F) 1 laptop. The category can be changed if the sets of appliances change among users. With this categorization, we now have a *profile* that can describe the building's basic occupancy.

*B. Data Collection*

We collect appliance energy consumption through a large-scale wireless sensor networks (WSNs). WSNs have been implemented in many different scenarios to facilitate the system estimation, conditioning, diagnosis **[2] [9] [10]**.

Specifically, a multi-channel DENT meter **[11]** is used to collect whole space real-time power consumption data. The DENT meter has 18 nodes, each one monitoring a subset of appliances, e.g. plug loads, lights, kitchenware etc.

Moreover, ACme sensors are used to collect real-time power consumption of each user in the space **[12]**. The data is handled using the sMAP protocol **[13]**. Each user has one ACme sensor installed to optimize the cost and experiment performance. The states of each appliance of each user are filtered out by the power dis-aggregation algorithm **[17]**.

*C. Power Dis-aggregation*

In power dis-aggregation, we *decode* the state ON/OFF of individual appliances from a *composite* power stream.

Mathematically, let $p_t$ be the observed composite power signal time series from $n$ appliances with $t$ from 1 to $T$. Let $S_t$ be the state vector of the $n$ appliances at step $t$. Our task is to infer $S_t$ from $p_t$. $S_t$ is a vector of $n$ binary variables, one for each appliance, i.e. $S_t \in \{0, 1\}$, in which 1 for ON. There are in total $2^n$ combinations of ON/OFF states.

Several models have been used to solve this problem, including Hidden Markov Model **[14]**, change detection **[15]**, sparse coding **[16]**. In this work, we use a method based on multiple hypothesis testing **[17]**.

### IV. MODEL FRAMEWORK

*A. Big Picture*

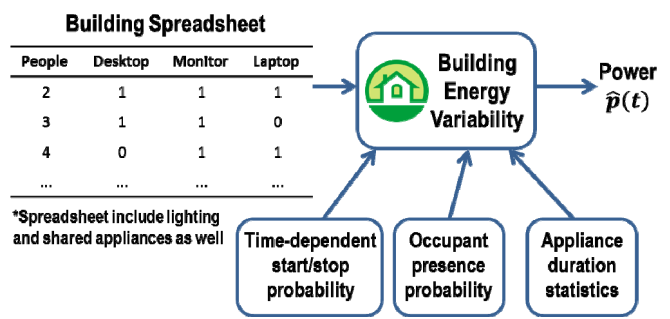

Fig. 2. Schematic of Model: Building End-Use Profile

In our work, we build a *Building Profile Model* (BPM) to estimate end-use energy profile. BPM generates variability through a black-box, and all we need for inputs are the occupants' information, and appliance categories. The BPM also takes several parameters include ON/OFF-probabilities, user presence probability, overnight probability, and/or appliance duration statistics, generalized from historical sensor recorded data and prior knowledge. The BPM has a great potential of model *re-use*, since in commercial building, in terms of scheduling, most of buildings in similar industries are essentially similar to each other in energy profile.

The schematic of the BPM is illustrated in **Fig 2**. In this section, we will start from the three widely used basic models, discuss their potential benefits in our scenario, and eventually arrive at our comprehensive BPM.

### B. Rate-of-Use Statistics Model

One basic model describing the appliance usage pattern utilizes the Rate-Of-Use (ROU) statistics, i.e. the portion of time that the appliance is in use given all samples. For example, in the 80 days of experiment, the monitor is ON at 12:00PM in 16 days, the ROU would be 0.2 at 12:00PM. The ROU is plotted for monitor, laptop and desktop in **Fig 3**. Strong daily pattern is observed. ROU provides good estimation of the average energy consumption, but little information about the variability of energy consumption.

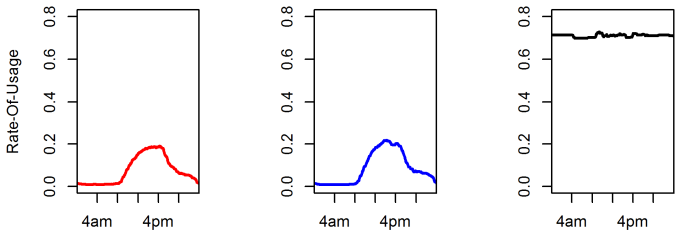

Fig. 3. ROU data of three types of appliances: monitor (left), laptop (middle) and desktop (right)

### C. ON/OFF-Probability Model

Another basic module is the ON/OFF-probability model **[7] [8]**, i.e. the probability of turning-ON/OFF at each time step. Compared to *Rate-of-Use statistics* model, this model can capture the appliances' stochasticity in both usage and power consumption **[2] [7] [8]**. Previously this model is built on TOU survey data, i.e. turning-ON of activities. Due to the low resolution of the data, we assume that the transition probabilities remain constant within preset time slots. For example, we can choose eight slots as "0~8AM", "8~9AM", "9~11:30AM", "11:30AM~1:30PM", "1:30~5PM", "5~7PM", "7~9:30PM" and "9:30PM~0AM", considering daily working and dining hours, morning rise-up and evening fall-down.

The ON-probability $P_{PW}$ is shown in **Fig 4**, for three office appliances: desktop, monitor and laptop. If we want to simulate appliance ON events in higher frequency, we use the probability $P_{PW}/N_{PW}$, in which $N_{PW}$ is the number of time steps in the time slots. For example, at slot "8~9AM", if we use 5 min interval simulation step, there are 12 points.

One concern about the time-slot-based ON probability model is that the probability inside each slot is not well captured. According to a simple Poisson model assuming independent events, within each time slot, the ON events are geometrically distributed. From the measurement, however, as shown in **Fig 5**, most of them do not follow the model.

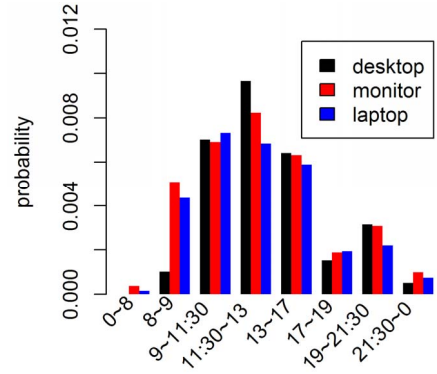

Fig. 4. Time-dependent ON probability of three types of appliances: desktop (black), monitor (red) and laptop (blue)

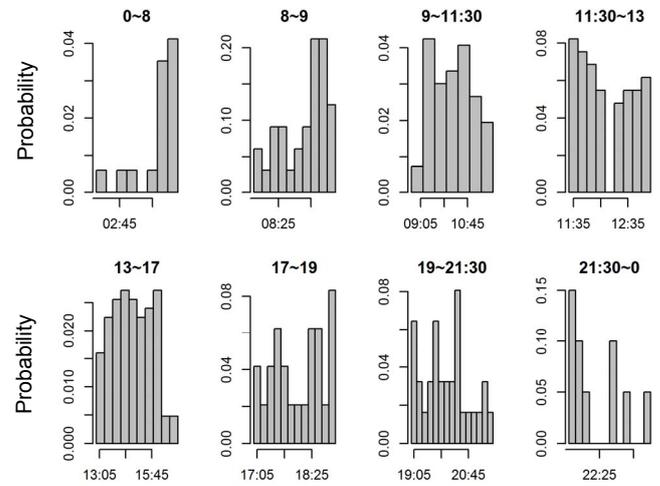

Fig. 5. Monitor ON probabilities inside each time slot for monitor

### D. Duration Statistics Model

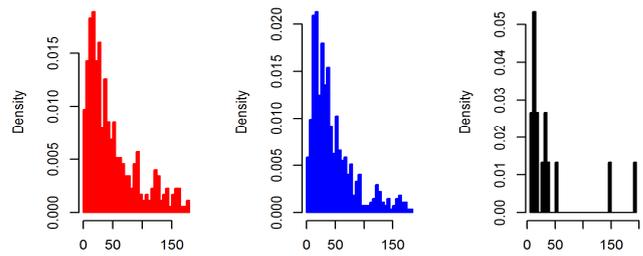

Fig. 6. Histogram of duration statistics of three types of appliances: monitor (left), laptop (middle) and desktop (right). X axis is in 5 minutes interval

Some previous work adopted ON probability and duration statistics model to generate stochastic sequences **[7] [8]**. Duration model measures the duration time of activities or appliances. We extracted the duration statistics of appliances from sensors after power dis-aggregation. The results are shown in **Fig 6** and seem to follow a Gamma distribution **[14]**.

A potential problem of this model is the limited capability to model the OFF events if the model is non-homogeneous. Usually, OFF events (e.g. around 6PM) needs to be enforced by an exponential smoothing function as:

$$D(t) \propto \frac{D_0(t)}{1+\exp\left(\frac{t-T_{\text{Off}}}{\lambda}\right)} \quad (1)$$

where $T_{\text{Off}}$ is the turning-off times, $\lambda$ is an exponential scale term. However, this enforcing function brings extra parameters into the model that will need to be estimated. Even if we create a time-dependent duration model $P_t(D)$, we are penalized by the increase in model complexity and over-fitting.

*E. Our Model*

In our model, we use an appliance-data-driven high resolution ON/OFF probability model. Firstly, we extract the probability that an appliance is present in some day, marked as $P_{\text{PRES}}$, as well as the probability that an appliance is ON overnight, marked as $P_{\text{INIT}}$, from the data. Secondly, we extract the empirical appliance ON/OFF probabilities $\hat{P}_t^{\text{ON}}$ and $\hat{P}_t^{\text{OFF}}$ from those days that $P_{\text{PRES}}=1$.

***Definition.1:** Empirical Appliance ON/OFF probabilities*

For the $i^{\text{th}}$ appliance at $t$, empirical ON/OFF probability $\hat{P}_{t,j}^{\text{ON}}$ & $\hat{P}_{t,j}^{\text{OFF}}$ from $M$ observed days is defined as:

$$\hat{P}_{t,j}^{\text{ON}} = \frac{\sum_{m=1}^{M} S_{t,j}^{(m)}\left(1-S_{t-1,j}^{(m)}\right)}{\sum_{m=1}^{M}\left(1-S_{t-1,j}^{(m)}\right)} \quad (2)$$

$$\hat{P}_{t,j}^{\text{OFF}} = \frac{\sum_{m=1}^{M} S_{t-1,j}^{(m)}\left(1-S_{t,j}^{(m)}\right)}{\sum_{m=1}^{M} S_{t-1,j}^{(m)}} \quad (3)$$

with which we can do Monte Carlo (MC) simulation of the $j^{\text{th}}$ appliance, and the simulated corresponding states in the $m^{\text{th}}$ run of the MC can be marked as $S_{t,j}^{(m)} \in \{0,1\}$, $t = 1, \ldots, T,$.

*F. Characteristics of the Model*

Firstly, since we have the wireless sensor network to directly collect the appliance power consumption data, we can build the model based on appliance information, instead of on activities as in other works, in which an often problematic activity-to-appliance transformation is needed **[4]**.

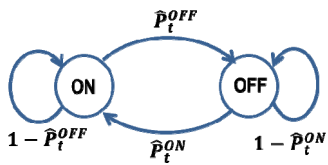

Fig. 7. Finite State Machine Intepretation of the ON/OFF Model

Secondly, both ON/OFF probabilities are included and are formulated in a Markov Chain framework, whereas duration statistics are not included. Therefore, we don't need to enforce the switching-off of appliances as in the duration model.

Thirdly, instead of an approximate time-slot based model, we use a more systematic non-homogeneous Markov Chain model for both ON & OFF probabilities. The model can be interpreted as a two state Finite State Machine (FSM) at each step (**Fig 7**) for each appliance, with $\hat{P}_t^{\text{ON}}$ & $\hat{P}_t^{\text{OFF}}$ as switching probabilities.

This model **(a)** can capture non-homogeneous stochasticity of appliance usage patterns and is easily extendable to analyze energy-saving policies and new techniques, as illustrated in other works about stochastic models; **(b)** is also statistically equivalent to the ROU model in estimating states, which means this method is essentially reasonable in end-use energy profile modeling. This can be shown in the theorem below:

***Theorem.1 Equivalence Theorem:*** If $\hat{S}_t$ is the time series of states from Monte Carlo simulation, then $\frac{1}{M}\sum_{m=1}^{M} S_t^m = \overline{S_t}$ is an unbiased estimator of $E\left[\hat{S}_t\right]$.

*Proof*: Let $\hat{S}_1, \ldots, \hat{S}_{t-1}, \hat{S}_t$ be the states at different time steps from MC simulation. Let us assume that the states are approximately following Markov Property, which means:

$$\Pr\left(\hat{S}_t \middle| \hat{S}_{t-1}, \ldots, \hat{S}_1\right) = \Pr\left(\hat{S}_t \middle| \hat{S}_{t-1}\right) \quad (4)$$

Then, by the chain rule of **[18]**, we have:

$$E\left[\hat{S}_t\right] = E\left[E\left[\hat{S}_t \middle| \hat{S}_{t-1}\right]\right] \quad (5)$$

Since we know:

$$\begin{aligned}E\left[\hat{S}_t \middle| \hat{S}_{t-1}\right] &= \Pr\left(\hat{S}_t = 1 \middle| \hat{S}_{t-1}\right) \\ &= \hat{P}_t^{\text{ON}}\left(1-\hat{S}_{t-1}\right) + \left(1-\hat{P}_t^{\text{OFF}}\right)\hat{S}_{t-1} \\ &= \hat{P}_t^{\text{ON}} + \left(1-\hat{P}_t^{\text{ON}}-\hat{P}_t^{\text{OFF}}\right)\hat{S}_{t-1}\end{aligned} \quad (6)$$

Combining **(5)** and **(6)** we have:

$$E\left[\hat{S}_t\right] = \hat{P}_t^{\text{ON}} + \left(1-\hat{P}_t^{\text{ON}}-\hat{P}_t^{\text{OFF}}\right)E\left[\hat{S}_{t-1}\right] \quad (7)$$

Let $G_t = 1-\hat{P}_t^{\text{ON}}-\hat{P}_t^{\text{OFF}}$, based on **(7)** we iteratively have:

$$E\left[\hat{S}_t\right] = \hat{P}_t^{\text{ON}} + \sum_{\tau=3}^{t}\prod_{i=\tau}^{t} G_i \hat{P}_{\tau-1}^{\text{ON}} + \prod_{i=2}^{t} G_i E\left[\hat{S}_1\right] \quad (8)$$

If we assume that $E\left[\hat{S}_1\right] = \overline{S_1}$ is known in advance, and put the expression of $\hat{P}_{t,j}^{\text{ON}}$ & $\hat{P}_{t,j}^{\text{OFF}}$ into **(8)**, we have this relationship:

$$\begin{aligned}&\prod_{i=3}^{t} G_i\left(\hat{P}_2^{\text{ON}} + \left(1-\hat{P}_2^{\text{ON}}-\hat{P}_2^{\text{OFF}}\right)\overline{S_1}\right) \\ &= \prod_{i=3}^{t} G_i\left(\frac{\overline{S_2}-\overline{S_1 S_2}}{1-\overline{S_1}} + \frac{\overline{S_1 S_2}-\overline{S_1}\cdot\overline{S_2}}{1-\overline{S_1}}\right) = \prod_{i=3}^{t} G_i \overline{S_2}\end{aligned} \quad (9)$$

Then recursively we can simply equation **(7)** as:

$$E\left[\hat{S}_t\right] = \frac{\overline{S_t} - \overline{S_{t-1}S_t}}{1 - \overline{S_{t-1}}} + \frac{\overline{S_{t-1}S_t} - \overline{S_{t-1}} \cdot \overline{S_t}}{1 - \overline{S_{t-1}}} = \overline{S_t} \quad (10)$$

Thus, we proved that the MC simulation is equivalent to the observed sample mean. Since $\hat{S}_t$ is a Bernoulli variable, $E[\hat{S}_t] = p_t$ is enough to evaluate all the statistical properties of $\hat{S}_t$. However, it should be noted that the proof of this theorem only works if the state distribution is consistent between the Monte Carlo simulation and the observations. If ON/OFF probability changes, the theorem never holds any more.

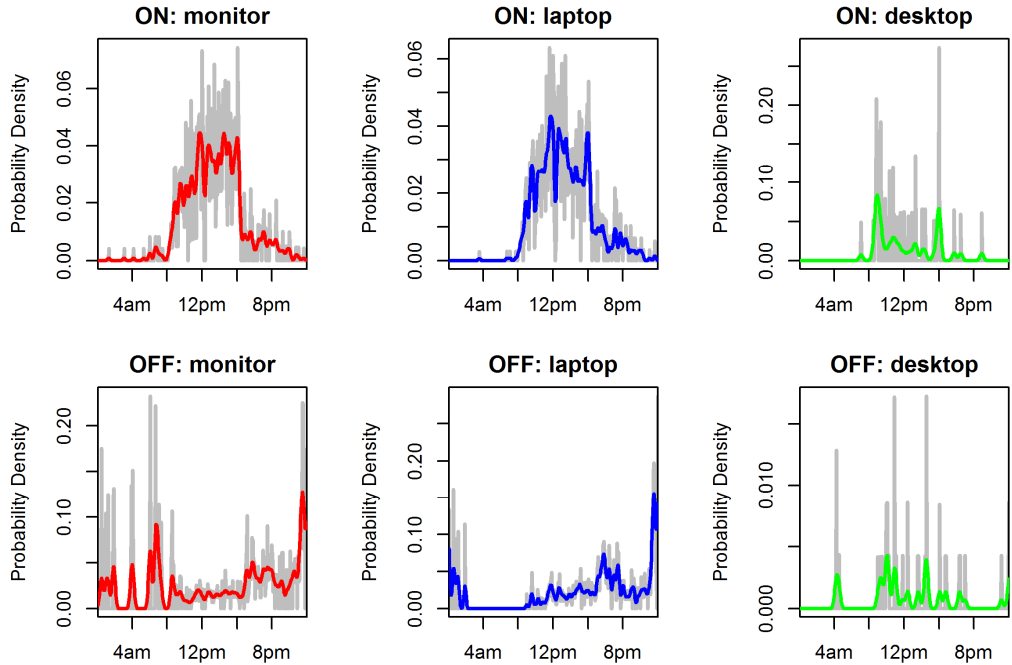

Fig. 8. ON/OFF Probability in 5 min interval for Monitor, Laptop, and Desktop. Gray lines: Measurement; Colored lines: Kernel smoothed

## V. RESULTS AND DISCUSSION

### A. ON/OFF Probability Model Estimation

As we estimate the ON/OFF probabilities, when the data points are sparse, we smooth the empirical probability function in **(2)** and **(3)** by a Kernel Smoother as below:

$$\tilde{P}_{t,j}^{\text{ON/OFF}} = \frac{\sum_{t=1}^{T} K(t, t_i) \hat{P}_{t_i,j}^{\text{ON/OFF}}}{\sum_{t=1}^{T} K(t, t_i)} \quad (11)$$

*Office Appliances*

The office appliances include *monitor, laptop, and desktop*. The estimated ON/OFF Probabilities for the three kinds of appliances are shown in **Fig 8**. It is observed that the starting (ON) probability peaks at early morning and decreases during the day, whereas stopping (OFF) probability peaks at late in the day. It should be noted that the data regarding to desktop is sparse and the ON/OFF probabilities contain more uncertainty. We only include *weekday*s in our study.

*Pathway/Room Lighting*

The lighting power consumption is a major contributor to building energy profile. In our test space in Cory 406 at UC Berkeley, we have pathway lighting and room lighting. Pathway lighting is shared in large working area and is more standard in schedule. Room lighting has motion sensor so that it is more adaptive to occupant behavior. The PowerScout data we collected contains the aggregated signal of lighting power in 7 rooms. For model simplicity, we assumes that the 7 rooms are the same. The result is shown in **Fig 9**.

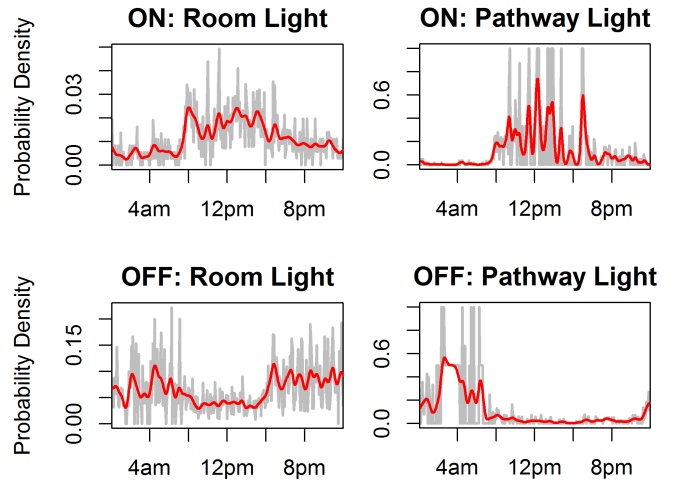

Fig. 9. ON/OFF Probability in 5 min interval for Room & pathway lighting. Gray lines: Measurement; Colored lines: Kernel smoothed

The pathway lighting has little overnight activity, and the estimation has more bias, since in **(3)**, $\overline{S_t}$ is zero for some *t*. We give those data point a probability of 0.5.

*Shared Appliances*

Shared appliances include a microwave, a water heater, a coffee maker, and a refrigerator. The water heater and refrigerator have strong periodic patterns, and are less dependent on occupants. The microwave and coffee maker shows spike-like patterns and the records of usage are sparse. The estimated probability densities for the Microwave are shown in **Fig 10**. Notice that the OFF probability is very high since the duration of each ON is usually very short, compared to our 5-minute estimation interval.

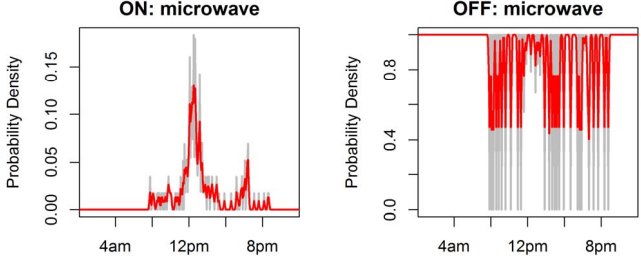

Fig. 10. ON/OFF Probability in 5 min interval for microwave. Gray lines: Measurement; Colored lines: Kernel smoothed

### B. Modeling of Cross-Correlation

In our experimental space, we have 11 monitors, 5 desktops, 14 laptops. Assuming that devices in the same category are the same, we can simulate each appliance independently and aggregate them. The mean of the aggregation, as a corollary of the ***Theorem 1***, is unbiased. The variance, however, could be underestimated. Cross-correlation among appliances needs to be addressed. In Monte Carlo simulation, cross-correlation between Bernoulli sequences is difficult. Instead, we propose a way to theoretically correct the variance estimation as follow.

Let $S_{t,i}$ be the state of $i^{th}$ single appliance, its variance $\text{var}(S_{t,i}) = \sigma_D^2$ we already know, $D$ = {desktop, monitor, laptop} is the appliance type, then the aggregated variance is:

$$\text{var}\left(\sum_{i=1}^{N} S_{t,i}\right) = \sum_{i=1}^{N} \sigma_{t,a(i)}^2 + \sum_{i \neq j} \text{cov}(S_{t,i}, S_{t,j}) \quad (12)$$

where $a(i)$ is the type of the $i^{th}$ appliance, and $\sum_{i=1}^{N} \sigma_{t,a(i)}^2 = \sum_{a \in D} \hat{\sigma}_{t,a}^2 N_a$, $N_a$ is the number of appliances in type $a$.

The term $\sum_{i=1}^{N} \sigma_{t,a(i)}^2$ is the uncorrelated variance, and the other term in RHS of **(12)** can be simplified as below:

$$\sum_{i \neq j} \text{cov}(S_{t,i}, S_{t,j}) = \sum_{a \in D} \hat{\sigma}_{t,a}^2 N_a (N_a - 1) \hat{\rho}_{a,a} + \sum_{a,b \in D} \hat{\sigma}_{t,a} \hat{\sigma}_{t,b} N_a N_b \hat{\rho}_{a,b} \quad (13)$$

where $\hat{\rho}_{a,a}$ is the average correlation within each types of appliance, and $\hat{\rho}_{a,b}$ is the average correlation between different types of appliances, both extracted from measurement.

### C. Monte Carlo Simulation

We use 10'000 runs of MC simulations. In each run, we follow the steps described here:

Firstly, we generate random sample with probability $P_{PRES}$, if the outcome is 0, the appliance is not present. If the outcome is 1, then we generate startup state $S_1 = \text{Ber}(P_{INIT})$.

Secondly, we simulate all the appliances of certain type and sum them up. After that, we correct the sample variance term with the cross-correlation terms as in **(12)** and **(13)**.

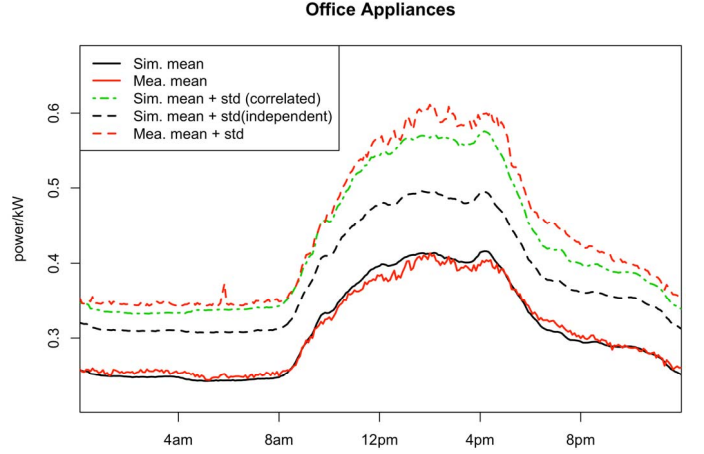

Fig. 11. MC Simulation for office appliances (11 monitors, 14 laptops and 5 desktops). The mean and standard deviation from measurements (red) and simulation (black) are shown. Theoretically corrected standard deviation is ploted in green to compare with the measurements.

The simulated end-use energy profiles are shown in **Fig 11**, **Fig 12**, and **Fig 13**, for office appliances, lighting, and representative shared appliances (we pick the microwave, other shared appliances work similarly), respectively. Both mean and standard deviation are extracted from MC simulation and only the upper bound of standard deviation is plotted since it is of more interest in early-stage demand estimation. Generally speaking, the model performs in all three categories of appliances. At the same time, we also have some interesting findings.

- Note that in **Fig 11**, cross correlation is shown to better capture the standard deviation level, which means that in the end-use profile modeling, the correlations among appliances have large impact on the overall variability.
- The standard deviation is poorly captured for microwave (as in **Fig 13**) and other appliances with spiking patterns, because of the sparse pattern. Variation-reduction techniques such as importance sampling or Markov Chain Monte Carlo (MCMC) could be used in the future to reduce the fluctuation.

It should be expected that, in a larger office building, when more appliances are present, our model can be more capable to capture the overnight patterns. Moreover, it should be noted that, when the building occupancy schematic changes, the only thing that needs to be tuned is the building *profile*. As long as we have a reasonable category of users, we can evaluate the building energy performance accordingly.

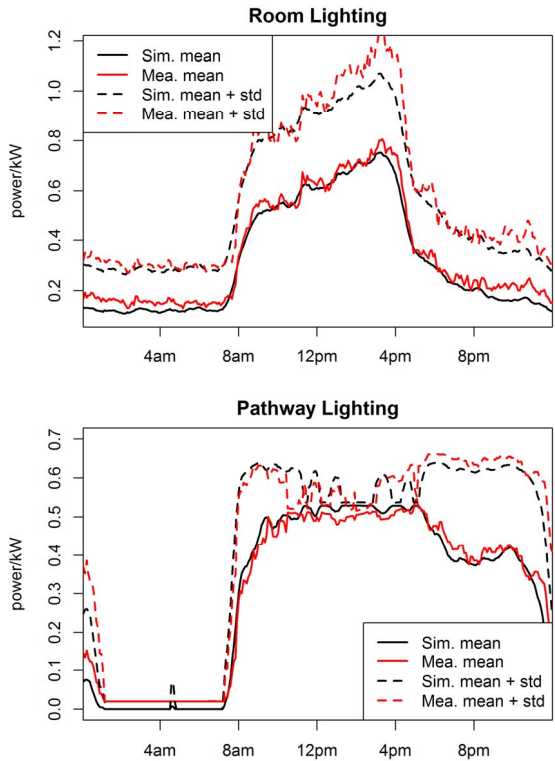

Fig. 12. MC Simulation Results for Room and Pathway Lighting.

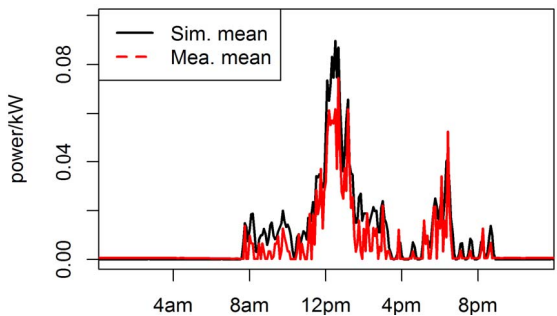

Fig. 13. MC Simulation Results for Microwave.

## VI. CONCLUSION

In this paper, the modeling of end-use energy profile is comprehensively investigated. The two categories *Top-down* and *Bottom-up* approaches are discussed and the latter is preferred because of the better integration with occupant-oriented information. Compared to the Time-Of-Use (TOU) data used in previous *Bottom-up* model, this work utilizes high frequency sampled data from wireless sensor network, and builds an appliance-data-driven end-use model. ON/OFF probabilities of appliances are extracted, and a theoretically unbiased FSM Markov Chain model is developed, with cross-correlation correction. The simulation results show the capability of the model to capture diversity and variability of building end-use energy profile, which can further help on the design of robust building power system.

ACKNOWLEDGMENT

This research is funded by the Republic of Singapore's National Research Foundation through a grant to the Berkeley Education Alliance for Research in Singapore (BEARS) for the Singapore-Berkeley Building Efficiency and Sustainability in the Tropics (SinBerBEST) Program. BEARS has been established by the University of California, Berkeley as a center for intellectual excellence in research and education in Singapore.